\def\eqn#1{\eqno(#1)}

\def\mi{\medskip\noindent}
\def\lbd{\{\hskip-5pt\{}
\def\rbd{\}\hskip-5pt\}}
\def\ladb{\left\langle\phantom{|^|_|}\hskip-12pt\right\langle}
\def\radb{\left\rangle\phantom{|^|_|}\hskip-12pt\right\rangle}
\def\lad{\langle\hskip-3pt\langle}
\def\rad{\rangle\hskip-3pt\rangle}
\def\la{\langle}
\def\ra{\rangle}
\def\bxi{\mbox{\boldmath$\xi$\unboldmath}}
\def\hxi{\hat{\mbox{\boldmath$\xi$\unboldmath}}\phantom{|}}
\documentclass[russian,12pt]{article}
\usepackage[T1]{fontenc}
\usepackage[cp866]{inputenc}
\usepackage{babel}
\begin{document}
{\center

\mi{\bf СЛАБО НЕЛИНЕЙНАЯ УСТОЙЧИВОСТЬ\\
МАГНИТОГИДРОДИНАМИЧЕСКИХ СИСТЕМ,\\
ИМЕЮЩИХ ЦЕНТР СИММЕТРИИ,\\
К ВОЗМУЩЕНИЯМ С БОЛЬШИМИ МАСШТАБАМИ}

\mi В.А.Желиговский

\mi Международный институт теории прогноза землетрясений\\
и математической геофизики РАН 

\mi Лаборатория общей аэродинамики, Институт механики МГУ\\


\mi{\it E-mail}: vlad@mitp.ru

}

\medskip Рассмотрена задача о слабо нелинейной устойчивости к возмущениям
с боль\-шими масштабами трехмерных магнитогидродинамических систем,
имеющих\break симметрию относительно центра. Предполагается, что в исследуемом
МГД состо\-янии отсутствуют большие пространственно-временн\'ые масштабы, и что эти
состояния устойчивы к возмущениям с таким же малым пространственным масш\-табом,
как в исследуемом состоянии. Выведенные с помощью асимптотических методов
уравнения для средних полей возмущений обобщают уравнения Навье-Стокса и
магнитной индукции. В них появляются оператор комбинированной вихревой
диффузии, вообще говоря, анизотропный и не обязательно отрицатель\-но
определенный, и дополнительные квадратичные члены, аналогичные адвектив\-ным.
Предложен метод экономичного вычисления коэффициентов вихревой диф\-фузии и
адвекции в уравнениях для средних полей.

\mi{\bf 1. Введение.}
Настоящая статья является непосредственным продолжением ра\-боты
Желиговского [2003], где была рассмотрена задача о линейной устойчивости
магнитогидродинамических (МГД) стационарных состояний. Предполагается, что характерный
прос\-транственный масштаб рассматриваемого стационарного состо\-яния существенно
меньше характерного масштаба возмущений. Тогда отношение этих масштабов
$\varepsilon$ -- малый параметр, и с помощью асимптотических методов
в указанной работе было построено решение соответствующей задачи
на собствен\-ные значения в виде степенных рядов по этому параметру. Было
показано, что главные члены разложений мод МГД возмущений
центрально-симметричных стационарных состояний и их инкрементов роста
являются соответственно собст\-венными векторами и собственными значениями
так называемого оператора ком\-бинированной вихревой (турбулентной) диффузии.
Это оператор в частных произ\-водных второго порядка, вообще говоря, анизотропный
и не обязательно знакооп\-ределенный. Если он имеет положительные
собственные значения, говорят о явлении отрицательной диффузии.
(Отрицательную диффузию в кинематичес\-ком динамо исследовали
Zheligovsky и др. [2001], Zheligovsky и Podvigina [2003] и Zheligovsky [2005],
в конвекции Рэлея-Бенара в присутствии магнитного поля -- Baptista и др. [2004].)

В настоящей работе рассмотрена дальнейшая эволюция возмущения в слабо
нелинейном режиме. МГД состояние ${\bf V,H},P$, нелинейная устойчивость
которого исследуется, удовлетворяет системе уравнений

$${\partial{\bf V}\over\partial t}=
\nu\Delta{\bf V}+{\bf V}\times(\nabla\times{\bf V})
+(\nabla\times{\bf H})\times{\bf H}-\nabla P+{\bf F},\eqn{1.1}$$
$${\partial{\bf H}\over\partial t}=
\eta\Delta{\bf H}+\nabla\times({\bf V}\times{\bf H})+{\bf J},\eqn{1.2}$$
$$\nabla\cdot{\bf V}=\nabla\cdot{\bf H}=0.\eqn{1.3}$$
Здесь ${\bf V(x},t)$ -- скорость потока проводящей жидкости,
${\bf H(x},t)$ -- магнитное поле, $P({\bf x},t)$ -- давление, $\nu$ и $\eta$ --
коэффициенты кинематической и магнитной молеку\-лярной диффузии, соответственно,
${\bf F(x},t)$ -- объемная сила, ${\bf J(x},t)$ отвечает нали\-чию в системе
распределения наложенных внешних токов. Предположение о стационарности
полей ${\bf V,H},P$ не делается.

Для системы в слабо нелинейном режиме считаем, что амплитуда возмущения
порядка $\varepsilon$. Возмущенное состояние \hbox{$\bf V+\varepsilon v$,}
${\bf H+\varepsilon h},P+\varepsilon p$ также удовлетворяет
системе (1), откуда профили возмущений ${\bf v,h},p$ (в дальнейшем эти поля
будем называть просто возмущениями) удовлетворяют уравнениям
$${\partial{\bf v}\over\partial t}=
\nu\Delta{\bf v}+{\bf v}\times(\nabla\times{\bf V})
+{\bf V}\times(\nabla\times{\bf v})+(\nabla\times{\bf h})\times{\bf H}
+(\nabla\times{\bf H})\times{\bf h}$$
$$+\varepsilon({\bf v}\times(\nabla\times{\bf v})
+(\nabla\times{\bf h})\times{\bf h})-\nabla p,\eqn{2.1}$$
$${\partial{\bf h}\over\partial t}=
\eta\Delta{\bf h}+\nabla\times({\bf v}\times{\bf H})
+\nabla\times({\bf V}\times{\bf h})
+\varepsilon\nabla\times({\bf v}\times{\bf h}),\eqn{2.2}$$
$$\nabla\cdot{\bf v}=\nabla\cdot{\bf h}=0.\eqn{2.3}$$

К системе уравнений (2) применяем методы теории осреднения уравнений в частных
производных [Bensoussan и др., 1978; Oleinik и др., 1992; Cioranescu и
Donato, 1999]. Вводим быстрые пространственные $\bf x$ и временн\'ую $t$ переменные
и соответствующие медленные переменные ${\bf X=\varepsilon x},T=\varepsilon^2t$
и представляем возму\-щение в виде формальных степенных рядов по
$\varepsilon$. Последовательно рассматриваем осредненную по быстрым переменным
и осциллирующую части полученных урав\-нений при каждой степени $\varepsilon^n$.
Эта процедура позволяет вывести (как условие разрешимости уравнений в быстрых
переменных при $n=2$) замкнутые нелиней\-ные уравнения в медленных переменных
для усредненного главного члена разло\-жения возмущения. В терминологии обзора
Newell и др.~[1993], эти уравнения (в дальнейшем будем называть их
уравнениями средних полей) описывают медленно модулированные конфигурации
течений (``order parameter equations for slowly modulated patterns"). Они обобщают
уравнения Навье-Стокса (1.1) и магнитной индукции (1.2). Вместо
операторов Лапласа в них появляется оператор комбини\-рованной вихревой
диффузии, как и в линейной задаче. (Таким образом, если имеет место эффект
отрицательной диффузии, то уравнения средних полей пере\-стают быть
параболическими.) Кроме того, в них присутствуют дополнительные
квадратичные члены, аналогичные адвективным. Тензоры вихревой
диффузии и вихревой адвекции определяются через решения
вспомогательных эллиптических (для стационарных конвективных
состояний) или параболических задач в быст\-рых переменных.

Уравнения средних полей для двумерных гидродинамических систем в отсутс\-твие
магнитного поля рассматривали в терминах функции тока Gama и др. [1994].

\mi{\bf 2. Предположения об исследуемом МГД состоянии ${\bf V,H},P$.}
В упомянутых выше статьях предполагалась периодичность полей ${\bf V,H},P$
по быстрым простран\-ственным переменным. Это условие избыточно,
и в настоящей работе оно не ставится. Считаем, что поля ${\bf V,H},P$,
определяющие исходное МГД состояние, устойчивость которого исследуется, гладки и
глобально ограничены, корректны все пространственно-временн\'ые усреднения,
проводимые в процессе вывода урав\-нений средних полей, и имеют решение
т.н. вспомогательные задачи, сформулиро\-ванные ниже. Как легко видеть, первые
два условия выполнены, если исходное состояние периодично по пространству
и стационарно или периодично по времени, или если оно квазипериодично
по пространству и времени, и энергетический спектр достаточно быстро затухает.
(Поле $f$ считается квазипериодичным по скалярной переменной $x$, если
$f(x)=\hat{f}(\alpha_1x,\,...\,,\alpha_mx)$, $\hat{f}$ периодично по
каждой своей переменной с одним и тем же периодом, и все
отношения констант $\alpha_{n_1}/\alpha_{n_2}$ при $n_1\ne n_2$ иррациональны.)
Имеется в виду усреднение по быстрым перемен\-ным:
$$\lad{\bf f}({\bf x},t,{\bf X},T)\rad\equiv\lim_{\phantom{|}\tau\to\infty}
\lim_{\phantom{|}\ell\to\infty}{1\over\tau\ell^3}\int_0^\tau\int_{-\ell/2}^{\ell/2}
\int_{-\ell/2}^{\ell/2}\int_{-\ell/2}^{\ell/2}
{\bf f}({\bf x},t,{\bf X},T)\,d{\bf x}\,dt$$
-- средняя, а $\lbd{\bf f}\rbd\equiv{\bf f}-\lad{\bf f}\rad$ -- осциллирующая
части поля $\bf f$.

Обозначим
$$\la f\ra\equiv\lim_{\ell\to\infty}{1\over\ell^3}
\int_{-\ell/2}^{\ell/2}\int_{-\ell/2}^{\ell/2}\int_{-\ell/2}^{\ell/2}
f({\bf x},...)\,d{\bf x}$$
-- пространственное (по быстрым переменным) среднее $f$.
Из цент\-ральной сим\-метричности исходного МГД состояния:
$${\bf V}(-{\bf x},t)=-{\bf V}({\bf x},t);\quad
{\bf H}(-{\bf x},t)=-{\bf H}({\bf x},t);\quad P(-{\bf x},t)=P({\bf x},t)$$
(считаем, что начало координат расположено в центре, относительно которого
симметрично МГД состояние) следует, что прост\-ранственные и, следовательно,
пространственно-временн\'ые средние этих полей равны нулю.

Предполагаем, что исходные состояния устойчивы к возмущениям с малым
пространственным масштабом. Операторы линеаризации исходных уравнений (2)
в окрестности исследуемого МГД состояния имеют вид
$${\cal L}^v({\bf w,g},q)\equiv-{\partial{\bf w}\over\partial t}
+\nu\Delta_{\bf x}{\bf w}+{\bf V}\times(\nabla_{\bf x}\times{\bf w})
+{\bf w}\times(\nabla_{\bf x}\times{\bf V})$$
$$+(\nabla_{\bf x}\times{\bf H})\times{\bf g}
+(\nabla_{\bf x}\times{\bf g})\times{\bf H}-\nabla_{\bf x}q,$$
$${\cal L}^h({\bf w,g})\equiv-{\partial{\bf g}\over\partial t}+\eta\Delta_{\bf x}{\bf g}
+\nabla_{\bf x}\times({\bf V}\times{\bf g}+{\bf w}\times{\bf H})$$

Пусть поля $\bf w,g$ и $q$ глобально ограничены вместе с их производными.
В силу (1.3) и поскольку $\bf V$ и $\bf H$ принадлежат этому классу,
\pagebreak
$$\la{\cal L}^v({\bf w,g},q)\ra=-{\partial\la{\bf w}\ra\over\partial t}
+\la{\bf V}\nabla_{\bf x}\cdot{\bf w}-{\bf H}\nabla_{\bf x}\cdot{\bf g}\ra;\eqn{3.1}$$
$$\la{\cal L}^h({\bf w},{\bf g})\ra=-{\partial\la{\bf g}\ra\over\partial t}\eqn{3.2}$$
$$\Rightarrow\quad\lad{\cal L}^v({\bf w,g},q)\rad=\lad{\bf V}\nabla_{\bf x}\cdot{\bf w}
-{\bf H}\nabla_{\bf x}\cdot{\bf g}\rad;\quad\lad{\cal L}^h({\bf w},{\bf g})\rad=0.\eqn{4}$$
Из (4) следует, что условия
$$\lad{\bf f}^v({\bf x},t)\rad=\lad{\bf f}^h({\bf x},t)\rad=0\eqn{5}$$
необходимы для существования решений системы уравнений
$${\cal L}^v({\bf w},{\bf g},q)={\bf f}^v,\quad{\cal L}^h({\bf w},{\bf g})={\bf f}^h,\eqn{6.1}$$
$$\nabla_{\bf x}\cdot{\bf w}=\nabla_{\bf x}\cdot{\bf g}=0\eqn{6.2}$$
из рассматриваемого класса. Вследствие (3) пространственное среднее решений
системы
$${\cal L}^v({\bf w},{\bf g},q)=0,\quad{\cal L}^h({\bf w},{\bf g})=0,\eqn{7.1}$$
$$\nabla_{\bf x}\cdot{\bf w}=\nabla_{\bf x}\cdot{\bf g}=0\eqn{7.2}$$
из этого класса функций сохраняется во времени, поэтому линейная устойчивость
исходного МГД состояния не асимптотическая. Будем считать,
что всякое решение (7) из указанного класса функций с нулевым пространственным
средним экспонен\-циально затухает во времени.

В дальнейшем предполагаем, что для произвольных глобально ограниченных вместе
с производными гладких соленоидальных полей ${\bf f}^v({\bf x},t)$,
${\bf f}^h({\bf x},t)$ с нулевым средним (6) имеет при заданных начальных
условиях единственное решение ${\bf w,g},q$ в указанном пространстве.
Это условие выполнено, например, для прост\-ранственно-периодических
стационарных или периодических по времени устой\-чивых к короткомасштабным
(не зависящим от медленных переменных) возмуще\-ниям состояний МГД систем
общего положения (малое возмущение полей $\bf F$ и $\bf J$ в системах, где оно
не выполнено, приводит либо к неустойчивости, либо к его выполнению).
Решение (6) может быть построено, как решение параболического уравнения
(однако поскольку область, которую занимает объем жидкости,
не компактна, нельзя гарантировать, что оно будет глобально ограничено
вместе с производными). Его единственность при данных начальных условиях
следует из сделанного выше
предположения о характере линейной устойчивости исходного МГД состояния.

\mi{\bf 3. Формальные асимптотические разложения.}
Решение задачи (2) ищем в виде степенных рядов
$${\bf v}=\sum_{n=0}^\infty{\bf v}_n({\bf x},t,{\bf X},T)\varepsilon^n,\qquad
{\bf h}=\sum_{n=0}^\infty{\bf h}_n({\bf x},t,{\bf X},T)\varepsilon^n,\eqn{8}$$
$$p=\sum_{n=0}^\infty p_n({\bf x},t,{\bf X},T)\varepsilon^n.\eqn{9}$$
Считаем, что в начальный момент времени все члены рядов (8) заданы.

Приравняем нулю коэффициенты рядов по степеням $\varepsilon$, полученных
подстанов\-кой рядов (8) в (2.3).
Выделяя среднюю и осциллирующую часть полученных уравнений, находим
$$\nabla_{\bf X}\cdot\lad{\bf v}_n\rad=\nabla_{\bf X}\cdot\lad{\bf h}_n\rad=0,\eqn{10.1}$$
$$\nabla_{\bf x}\cdot\lbd{\bf v}_n\rbd+\nabla_{\bf X}\cdot\lbd{\bf v}_{n-1}\rbd=
\nabla_{\bf x}\cdot\lbd{\bf h}_n\rbd+\nabla_{\bf X}\cdot\lbd{\bf h}_{n-1}\rbd=0\eqn{10.2}$$
при всех $n\ge0$. Здесь и далее в дифференциальных операторах с индексами $\bf x$ и
$\bf X$ дифференцирование производится по быстрым и медленным пространствен\-ным
переменным, соответственно (все члены рядов с индексом $n<0$
по определе\-нию равны 0).

Подставив (8) и (9) в уравнения (2.1) и (2.2), преобразуем последние
к виду равенств рядов по степеням $\varepsilon$. Приравнивая коэффициенты
этих рядов, получаем рекуррентную систему уравнений, которую последовательно
решаем совместно с условиями (10),
выделяя среднюю и осциллирующую часть каждого уравнения.

\mi{\bf 4. Уравнения порядка $\varepsilon^0$.}
Из главных членов рядов (2.1) и (2.2) получаем уравнения
$${\cal L}^v(\lbd{\bf v}_0\rbd,\lbd{\bf h}_0\rbd,\lbd p_0\rbd)+\lad{\bf v}_0\rad\times(\nabla_{\bf x}\times{\bf V})
+(\nabla_{\bf x}\times{\bf H})\times\lad{\bf h}_0\rad=0,\eqn{11.1}$$
$${\cal L}^h(\lbd{\bf v}_0\rbd,\lbd{\bf h}_0\rbd)+(\lad{\bf h}_0\rad\cdot\nabla_{\bf x}){\bf V}
-(\lad{\bf v}_0\rad\cdot\nabla_{\bf x}){\bf H}=0.\eqn{11.2}$$
Поскольку средние $\lad{\bf v}_0\rad$ и $\lad{\bf h}_0\rad$ не зависят от
быстрых переменных, а в операторах ${\cal L}^v$ и ${\cal L}^h$ дифференцирование проводится
только по быстрым переменным, в силу линейности (11) эти уравнения
имеют решения следующей структуры:
$$\lbd{\bf v}_0\rbd=\bxi_0^v+\sum_{k=1}^3({\bf S}^{v,v}_k\lad{\bf v}^k_0\rad
+{\bf S}^{h,v}_k\lad{\bf h}^k_0\rad),\eqn{12.1}$$
$$\lbd{\bf h}_0\rbd=\bxi_0^h+\sum_{k=1}^3({\bf S}^{v,h}_k\lad{\bf v}^k_0\rad
+{\bf S}^{h,h}_k\lad{\bf h}^k_0\rad),\eqn{12.2}$$
$$\lbd p_0\rbd=\xi_0^p+\sum_{k=1}^3(S^{v,p}_k\lad{\bf v}^k_0\rad+S^{h,p}_k\lad{\bf h}^k_0\rad),\eqn{12.3}$$
где функции ${\bf S}^{\cdot,\cdot}_k({\bf x},t)$ являются решениями
{\it вспомогательных задач первого типа}
$${\cal L}^v({\bf S}^{v,v}_k,{\bf S}^{v,h}_k,S^{v,p}_k)=
-{\bf e}_k\times(\nabla_{\bf x}\times{\bf V}),\eqn{13.1}$$
$$\nabla_{\bf x}\cdot{\bf S}^{v,v}_k=0,\eqn{13.2}$$
$${\cal L}^h({\bf S}^{v,v}_k,{\bf S}^{v,h}_k)=
{\partial{\bf H}\over\partial x_k}\eqn{13.3}$$
$$\nabla_{\bf x}\cdot{\bf S}^{v,h}_k=0;\eqn{13.4}$$
$${\cal L}^v({\bf S}^{h,v}_k,{\bf S}^{h,h}_k,S^{h,p}_k)=
{\bf e}_k\times(\nabla_{\bf x}\times{\bf H}),\eqn{14.1}$$
$$\nabla_{\bf x}\cdot{\bf S}^{h,v}_k=0,\eqn{14.2}$$
$${\cal L}^h({\bf S}^{h,v}_k,{\bf S}^{h,h}_k)=
-{\partial{\bf V}\over\partial x_k},\eqn{14.3}$$
$$\nabla_{\bf x}\cdot{\bf S}^{h,h}_k=0,\eqn{14.4}$$
\pagebreak
а $\bxi_0({\bf x},t,{\bf X},T)$ удовлетворяют
$${\cal L}^v(\bxi_0^v,\bxi_0^h,\xi_0^p)={\cal L}^h(\bxi_0^v,\bxi_0^h)=0,\quad
\nabla_{\bf x}\cdot\bxi_0^v=\nabla_{\bf x}\cdot\bxi_0^h=0,\quad\la\bxi_0\ra=0.\eqn{15}$$
Здесь ${\bf e}_k$ -- единичный вектор вдоль оси координат $x_k$, верхний
индекс $k$ нумерует компоненты вектора:
$$\lad{\bf v}_0\rad=\sum_{k=1}^3\lad{\bf v}^k_0\rad{\bf e}_k,\qquad
\lad{\bf h}_0\rad=\sum_{k=1}^3\lad{\bf h}^k_0\rad{\bf e}_k.$$
Условия (13.2), (13.4), (14.2) и (14.4) необходимы для выполнения (10.2) при $n=0$.

В качестве начальных условий для задач (13) и (14) выберем некоторые
глобально ограниченные вместе с производными гладкие соленоидальные поля,
антисимметричные относительно центра, с нулевым
пространственным средним. Тогда (13) и (14) имеют единственные
решения согласно предположению о разре\-шимости задач (6) (условие разрешимости
(5) для них выполнено в силу глобаль\-ной ограниченности полей $\bf V$ и~$\bf H$).
Взяв дивергенцию уравнений (13.3) и (14.3), находим, что для выполнения (13.4)
и (14.4) при $t>0$ необходимо и достаточно потребовать соленоидальность
${\bf S}^{\cdot,h}_k$ при $t=0$.
Пространственные средние правых частей уравнений (13.1), (13.3), (14.1) и (14.3)
равны нулю, поэтому требование $\la{\bf S}^{\cdot,\cdot}_k\ra=0$ при $t=0$ влечет
тогда $\lad{\bf S}^{\cdot,\cdot}_k\rad=0$ в силу (3). Аналогично, выполнение
условий $\nabla_{\bf x}\cdot\bxi_0^h=0$ и
$\la\bxi_0\ra=0$ достаточно требовать только при $t=0$. Вследствие центральной
симметричности исходного МГД состояния пространства централь\-но-симметричных
и центрально-антисимметричных полей инвариантны для опе\-ратора ${\cal L}=({\cal L}^v,{\cal L}^h)$,
поэтому решения ${\bf S}^{\cdot,\cdot}_k$ соответствуют центрально-антисим\-метричным состояниям:
$${\bf S}^{\cdot,v}_k(-{\bf x},t)={\bf S}^{\cdot,v}_k({\bf x},t),\quad
{\bf S}^{\cdot,h}_k(-{\bf x},t)={\bf S}^{\cdot,h}_k({\bf x},t),$$
$$S^{\cdot,p}_k(-{\bf x},t)=-S^{\cdot,p}_k({\bf x},t).$$

Начальные условия для задачи (15) выбираем из условий
$${\bf v}_0=\bxi_0^v+\sum_{k=1}^3\left({\bf S}^{v,v}_k\lad{\bf v}^k_0\rad
+{\bf S}^{h,v}_k\lad{\bf h}^k_0\rad\right)+\lad{\bf v}_0\rad,\eqn{16.1}$$
$${\bf h}_0=\bxi_0^h+\sum_{k=1}^3\left({\bf S}^{v,h}_k\lad{\bf v}^k_0\rad
+{\bf S}^{h,h}_k\lad{\bf h}^k_0\rad\right)+\lad{\bf h}_0\rad\eqn{16.2}$$
при $t=0$. Усредняя (16) по быстрой пространственной переменной, находим
$$\lad{\bf v}_0\rad|_{T=0}=\la{\bf v}_0\ra|_{t=0},\quad
\lad{\bf h}_0\rad|_{T=0}=\la{\bf h}_0\ra|_{t=0},$$
а из осциллирующих по быстрой пространственной переменной части условий (16)
находим начальные условия для задачи (15). Изменение начальных условий для
${\bf S}^{\cdot,\cdot}_k$ в рассматриваемом классе начальных условий
компенсируется соответст\-вующим изменением начальных условий для $\bxi^\cdot_0$,
однако, как будет показано ниже, эта неоднозначность не влияет на вид уравнений
для средних полей, посколь\-ку эти изменения ${\bf S}^{\cdot,\cdot}_k$ и $\bxi_0$
экспоненциально затухают во времени (т.к. они являются решениями задачи (7)
с нулевыми пространственными средними). В случае, если исходное МГД
состояние стационарно или периодично по быстрому времени,
для удобства вычисления пространственно-временн\'ых средних
естест\-венно потребо\-вать, чтобы функции ${\bf S}^{\cdot,\cdot}_k$ были, соответственно,
стационарными или периодичны\-ми по времени решениями задач (13) и (14).
Существование таких стационарных решений при рассматриваемых условиях доказано
Желиговским [2003] ( (13) и (14) сводятся тогда к первой вспомогательной задаче,
рассмотренной в указанной статье); периодический по времени случай
рассматривается аналогично (см. Zheligovsky и Podvigina [2003]).
Тогда $\bxi_0$ имеет смысл затухающих переходных про\-цессов, приводящих
к установившемуся (в быстром времени) режиму.

\mi{\bf 5. Уравнения порядка $\varepsilon^1$.}
Уравнения, полученные из членов рядов (2.1) и (2.2) порядка $\varepsilon$,
имеют вид
$${\cal L}^v(\lbd{\bf v}_1\rbd,\lbd{\bf h}_1\rbd,\lbd p_1\rbd)
+\lad{\bf v}_1\rad\times(\nabla_{\bf x}\times{\bf V})
+(\nabla_{\bf x}\times{\bf H})\times\lad{\bf h}_1\rad$$
$$+2\nu(\nabla_{\bf x}\cdot\nabla_{\bf X})\lbd{\bf v}_0\rbd
+{\bf V}\times(\nabla_{\bf X}\times{\bf v}_0)
+(\nabla_{\bf X}\times{\bf h}_0)\times{\bf H}$$
$$+{\bf v}_0\times(\nabla_{\bf x}\times\lbd{\bf v}_0\rbd)
+(\nabla_{\bf x}\times\lbd{\bf h}_0\rbd)\times{\bf h}_0-\nabla_{\bf X}p_0=0,\eqn{17.1}$$
$${\cal L}^h(\lbd{\bf v}_1\rbd,\lbd{\bf h}_1\rbd)+(\lad{\bf h}_1\rad\cdot\nabla_{\bf x}){\bf V}
-(\lad{\bf v}_1\rad\cdot\nabla_{\bf x}){\bf H}
+2\eta(\nabla_{\bf x}\cdot\nabla_{\bf X})\lbd{\bf h}_0\rbd$$
$$+\nabla_{\bf X}\times({\bf v}_0\times{\bf H}+{\bf V}\times{\bf h}_0)
+\nabla_{\bf x}\times({\bf v}_0\times{\bf h}_0)=0\eqn{17.2}$$
(здесь использованы соотношения (10) при $n=1$). В~силу линейности
этих уравнений они имеют решения следующей структуры:
$$\lbd{\bf v}_1\rbd=\bxi_1^v+\sum_{k=1}^3\left({\bf S}^{v,v}_k\lad{\bf v}^k_1\rad
+{\bf S}^{h,v}_k\lad{\bf h}^k_1\rad+\sum_{m=1}^3\left(
{\bf G}^{v,v}_{m,k}{\partial\lad{\bf v}^k_0\rad\over\partial X_m}
+{\bf G}^{h,v}_{m,k}{\partial\lad{\bf h}^k_0\rad\over\partial X_m}\right.\right.$$
$$\left.\left.\phantom{|^|\over|_|}\!\!\!
+{\bf Q}^{vv,v}_{m,k}\lad{\bf v}^k_0\rad\lad{\bf v}^m_0\rad
+{\bf Q}^{vh,v}_{m,k}\lad{\bf v}^k_0\rad\lad{\bf h}^m_0\rad
+{\bf Q}^{hh,v}_{m,k}\lad{\bf h}^k_0\rad\lad{\bf h}^m_0\rad\right)\right),\eqn{18.1}$$
$$\lbd{\bf h}_1\rbd=\bxi_1^h+\sum_{k=1}^3\left({\bf S}^{v,h}_k\lad{\bf v}^k_1\rad
+{\bf S}^{h,h}_k\lad{\bf h}^k_1\rad+\sum_{m=1}^3\left(
{\bf G}^{v,h}_{m,k}{\partial\lad{\bf v}^k_0\rad\over\partial X_m}
+{\bf G}^{h,h}_{m,k}{\partial\lad{\bf h}^k_0\rad\over\partial X_m}\right.\right.$$
$$\left.\left.\phantom{|^|\over|_|}\!\!\!
+{\bf Q}^{vv,h}_{m,k}\lad{\bf v}^k_0\rad\lad{\bf v}^m_0\rad
+{\bf Q}^{vh,h}_{m,k}\lad{\bf v}^k_0\rad\lad{\bf h}^m_0\rad
+{\bf Q}^{hh,h}_{m,k}\lad{\bf h}^k_0\rad\lad{\bf h}^m_0\rad\right)\right),\eqn{18.2}$$
$$\lbd p_1\rbd=\xi_1^p+\sum_{k=1}^3\left(S^{v,p}_k\lad{\bf v}^k_1\rad
+S^{h,p}_k\lad{\bf h}^k_1\rad+\sum_{m=1}^3\left(
G^{v,p}_{m,k}{\partial\lad{\bf v}^k_0\rad\over\partial X_m}
+G^{h,p}_{m,k}{\partial\lad{\bf h}^k_0\rad\over\partial X_m}\right.\right.$$
$$\left.\left.\phantom{|^|\over|_|}\!\!\!
+Q^{vv,p}_{m,k}\lad{\bf v}^k_0\rad\lad{\bf v}^m_0\rad
+Q^{vh,p}_{m,k}\lad{\bf v}^k_0\rad\lad{\bf h}^m_0\rad
+Q^{hh,p}_{m,k}\lad{\bf h}^k_0\rad\lad{\bf h}^m_0\rad\right)\right),\eqn{18.3}$$
где функции $\bf G$ являются решениями {\it вспомогательных задач второго типа}:
\pagebreak
$${\cal L}^v({\bf G}^{v,v}_{m,k},{\bf G}^{v,h}_{m,k},G^{v,p}_{m,k})=
-2\nu{\partial{\bf S}^{v,v}_k\over\partial x_m}-{\bf V}^k{\bf e}_m
+{\bf V}^m{\bf e}_k-({\bf V}\cdot{\bf S}^{v,v}_k){\bf e}_m$$
$$+{\bf V}^m{\bf S}^{v,v}_k+{\bf e}_mS^{v,p}_k
+({\bf H}\cdot{\bf S}^{v,h}_k){\bf e}_m-{\bf H}^m{\bf S}^{v,h}_k\eqn{19.1}$$
$$\nabla_{\bf x}\cdot{\bf G}^{v,v}_{m,k}=-({\bf S}^{v,v}_k)^m,\eqn{19.2}$$
$${\cal L}^h({\bf G}^{v,v}_{m,k},{\bf G}^{v,h}_{m,k})=
-2\eta{\partial{\bf S}^{v,h}_k\over\partial x_m}
-{\bf e}_m\times\left({\bf V}\times{\bf S}^{v,h}_k
+({\bf S}^{v,v}_k+{\bf e}_k)\times{\bf H}\right);\eqn{19.3}$$
$$\nabla_{\bf x}\cdot{\bf G}^{v,h}_{m,k}=-({\bf S}^{v,h}_k)^m,\eqn{19.4}$$
$${\cal L}^v({\bf G}^{h,v}_{m,k},{\bf G}^{h,h}_{m,k},G^{h,p}_{m,k})=
-2\nu{\partial{\bf S}^{h,v}_k\over\partial x_m}+{\bf H}^k{\bf e}_m
-{\bf H}^m{\bf e}_k-({\bf V}\cdot{\bf S}^{h,v}_k){\bf e}_m$$
$$+{\bf V}^m{\bf S}^{h,v}_k+{\bf e}_mS^{h,p}_k
+({\bf H}\cdot{\bf S}^{h,h}_k){\bf e}_m-{\bf H}^m{\bf S}^{h,h}_k\eqn{20.1}$$
$$\nabla_{\bf x}\cdot{\bf G}^{h,v}_{m,k}=-({\bf S}^{h,v}_k)^m,\eqn{20.2}$$
$${\cal L}^h({\bf G}^{h,v}_{m,k},{\bf G}^{h,h}_{m,k})=
-2\eta{\partial{\bf S}^{h,h}_k\over\partial x_m}
-{\bf e}_m\times\left({\bf V}\times({\bf S}^{h,h}_k+{\bf e}_k)
+{\bf S}^{h,v}_k\times{\bf H}\right);\eqn{20.3}$$
$$\nabla_{\bf x}\cdot{\bf G}^{h,h}_{m,k}=-({\bf S}^{h,h}_k)^m,\eqn{20.4}$$
функции $\bf Q$ -- решениями {\it вспомогательных задач третьего типа}:
$${\cal L}^v({\bf Q}^{vv,v}_{m,k},{\bf Q}^{vv,h}_{m,k},Q^{vv,p}_{m,k})=
-({\bf S}^{v,v}_k+{\bf e}_k)\times(\nabla_{\bf x}\times{\bf S}^{v,v}_m)
-(\nabla_{\bf x}\times{\bf S}^{v,h}_k)\times{\bf S}^{v,h}_m,\eqn{21.1}$$
$${\cal L}^h({\bf Q}^{vv,v}_{m,k},{\bf Q}^{vv,h}_{m,k})=-\nabla_{\bf x}\times
(({\bf S}^{v,v}_k+{\bf e}_k)\times{\bf S}^{v,h}_m)\eqn{21.2}$$
$${\cal L}^v({\bf Q}^{vh,v}_{m,k},{\bf Q}^{vh,h}_{m,k},Q^{vh,p}_{m,k})=
-({\bf S}^{v,v}_k+{\bf e}_k)\times(\nabla_{\bf x}\times{\bf S}^{h,v}_m)
-{\bf S}^{h,v}_m\times(\nabla_{\bf x}\times{\bf S}^{v,v}_k)$$
$$-(\nabla_{\bf x}\times{\bf S}^{v,h}_k)\times({\bf S}^{h,h}_m+{\bf e}_m)
-(\nabla_{\bf x}\times{\bf S}^{h,h}_m)\times{\bf S}^{v,h}_k,\eqn{22.1}$$
$${\cal L}^h({\bf Q}^{vh,v}_{m,k},{\bf Q}^{vh,h}_{m,k})=-\nabla_{\bf x}\times
(({\bf S}^{v,v}_k+{\bf e}_k)\times({\bf S}^{h,h}_m+{\bf e}_m)+
{\bf S}^{h,v}_m\times{\bf S}^{v,h}_k);\eqn{22.2}$$
$${\cal L}^v({\bf Q}^{hh,v}_{m,k},{\bf Q}^{hh,h}_{m,k},Q^{hh,p}_{m,k})=
-{\bf S}^{h,v}_k\times(\nabla_{\bf x}\times{\bf S}^{h,v}_m)
-(\nabla_{\bf x}\times{\bf S}^{h,h}_k)\times({\bf S}^{h,h}_m+{\bf e}_m),\eqn{23.1}$$
$${\cal L}^h({\bf Q}^{hh,v}_{m,k},{\bf Q}^{hh,h}_{m,k})=-\nabla_{\bf x}\times
({\bf S}^{h,v}_k\times({\bf S}^{h,h}_m+{\bf e}_m))\eqn{23.2}$$
$$\nabla_{\bf x}\cdot{\bf Q}^{\cdot\cdot,\cdot}_{m,k}=0,\eqn{24}$$
а функции $\bxi_1$ -- решениями задач
$${\cal L}^v(\bxi_1^v,\bxi_1^h,\xi_1^p)=-2\nu(\nabla_{\bf x}\cdot\nabla_{\bf X})\bxi_0^v
-{\bf V}\times(\nabla_{\bf X}\times\bxi_0^v)
-(\nabla_{\bf X}\times\bxi_0^h)\times{\bf H}
-\bxi_0^v\times(\nabla_{\bf x}\times{\bf v}_0)$$
$$-(\nabla_{\bf x}\times{\bf h}_0)\times\bxi_0^h
-({\bf v}_0-\bxi_0^v)\times(\nabla_{\bf x}\times\bxi_0^v)
-(\nabla_{\bf x}\times\bxi_0^h)\times({\bf h}_0-\bxi_0^h)
+\nabla_{\bf X}\xi_0^p,\eqn{25.1}$$
$$\nabla_{\bf x}\cdot\bxi_1^v+\nabla_{\bf X}\cdot\bxi_0^v=0,\eqn{25.2}$$
$${\cal L}^h(\bxi_1^v,\bxi_1^h)=-2\eta(\nabla_{\bf x}\cdot\nabla_{\bf X})\bxi_0^h
-({\bf H}\cdot\nabla_{\bf X})\bxi_0^v+({\bf V}\cdot\nabla_{\bf X})\bxi_0^h
-{\bf V}\nabla_{\bf X}\cdot\bxi_0^h+{\bf H}\nabla_{\bf X}\cdot\bxi_0^v,$$
$$-(\bxi_0^h\cdot\nabla_{\bf x}){\bf v}_0+(\bxi_0^v\cdot\nabla_{\bf x}){\bf h}_0
-(({\bf h}_0-\bxi_0^h)\cdot\nabla_{\bf x})\bxi_0^v
+(({\bf v}_0-\bxi_0^v)\cdot\nabla_{\bf x})\bxi_0^h,\eqn{25.3}$$
$$\nabla_{\bf x}\cdot\bxi_1^h+\nabla_{\bf X}\cdot\bxi_0^h=0.\eqn{25.4}$$

Условия (19.2), (19.4), (20.2), (20.4), (25.2), (25.4) и (24) гарантируют
выполне\-ние (10.2) при $n=1$. Взяв дивергенцию уравнений (19.3) и (20.3)
и комбинируя полученные равенства с (13.3) и (14.3), соответственно, получаем
равенства
$$\left(-{\partial\over\partial t}+\eta\Delta_{\bf x}\right)
\left(\nabla_{\bf x}\cdot{\bf G}^{\cdot,h}_{m,k}+({\bf S}^{\cdot,h}_k)^m\right)=0,$$
откуда (19.4) и (20.4) выполнены при $t>0$ тогда и только тогда, когда
они выполнены при $t=0$. Из дивергенций уравнений (21.2), (22.2) и (23.2)
находим, что поля ${\bf Q}^{h,h}_{m,k}$ соленоидальны при $t>0$ если и только
если они соленоидальны при $t=0$. Взяв дивергенцию уравнения (25.3) по быстрым
переменным, находим, используя равенство ${\cal L}^h(\bxi_0^v,\bxi_0^h)=0$ (15),
$$\left(-{\partial\over\partial t}+\eta\Delta_{\bf x}\right)(\nabla_{\bf x}\cdot\bxi_1^h)
=-\nabla_{\bf x}\cdot\left(\nabla_{\bf X}\times(\bxi_0^v\times{\bf H}+{\bf V}\times\bxi_0^h)\right)$$
$$=\nabla_{\bf X}\cdot\left(\nabla_{\bf x}\times(\bxi_0^v\times{\bf H}+{\bf V}\times\bxi_0^h)\right)
=-\left(-{\partial\over\partial t}+\eta\Delta_{\bf x}\right)(\nabla_{\bf X}\cdot\bxi_0^h),$$
откуда (25.4) выполнено при $t>0$ тогда и только тогда, когда оно
выполнено при $t=0$.

В качестве начальных условий для задач (19)-(24) выберем некоторые
глобаль\-но ограниченные вместе с производными гладкие поля, удовлетворяющие
(19.2), (19.4), (20.2), (20.4) и (24), антисимметричные относительно центра
(пространст\-венные средние которых следовательно равны нулю).
Начальными условиями для $\bxi_1$ могут служить любые глобально ограниченные
вместе с производными гладкие поля, удовлетворяющие (25.2) и (25.4). Интегрируя
(25.1) по быстрому времени и усредняя результат с учетом равенств (3),
(25.2) и $\lad\bxi_1^v\rad=0$ (см. (18)), находим
$$\la\bxi_1^v\ra|_{t=0}=\ladb\int_0^t\left({\bf V}\nabla_{\bf X}\cdot\bxi_0^v
-{\bf H}\nabla_{\bf X}\cdot\bxi_0^h\right.$$
$$\left.-{\bf V}\times(\nabla_{\bf X}\times\bxi_0^v)
-(\nabla_{\bf X}\times\bxi_0^h)\times{\bf H}+\nabla_{\bf X}\xi_0^p\right)\,dt\radb;$$
аналогично из (25.4)
$$\la\bxi_1^h\ra|_{t=0}=-\ladb\int_0^t\nabla_{\bf X}\times\left(
{\bf V}\times\bxi_0^h+\bxi_0^v\times{\bf H}\right)\,dt\radb.$$
Поскольку пространственное среднее каждой суммы по $k$ в (18) равно 0,
$$\la{\bf v}_1\ra=\lad{\bf v}_1\rad+\la\bxi_1^v\ra;\quad
\la{\bf h}_1\ra=\lad{\bf h}_1\rad+\la\bxi_1^h\ra.$$
Отсюда, зная ${\bf v}_1|_{t=0}$ и ${\bf h}_1|_{t=0}$, находим
$\lad{\bf v}_1\rad|_{T=0}$ и $\lad{\bf h}_1\rad|_{T=0}$, а затем из (18.1)
и (18.2), где $\lbd{\bf v}_1\rbd={\bf v}_1-\lad{\bf v}_1\rad$ и
$\lbd{\bf h}_1\rbd={\bf h}_1-\lad{\bf h}_1\rad$,
находим начальные условия для задачи (25). Изменение начальных условий для
${\bf G}^{\cdot,\cdot}_{m,k}$ и ${\bf Q}^{\cdot\cdot,\cdot}_{m,k}$
в рассматриваемом классе начальных условий компенсируется соответствующим
изменением началь\-ных условий для $\bxi^\cdot_1$, однако, как будет ясно из
дальнейшего, эта неоднозначность не влияет на вид уравнений
для средних полей, поскольку вызванные этим изменения полей
${\bf G}^{\cdot,\cdot}_{m,k}$ и ${\bf Q}^{\cdot\cdot,\cdot}_{m,k}$
экспоненциально затухают во времени (т.к. они являются решениями задачи (7)
с нулевыми пространственными средними), и поэтому не дают вклад в
пространственно-временн\'ые средние, определяющие коэффициенты в новых членах,
появляющиеся в уравнениях для средних полей. В случае, если исходное МГД
состояние стационарно или периодично по (быстро\-му) времени,
для удобства вычисления пространственно-временн\'ых средних
ес\-тест\-венно потребовать, чтобы функции ${\bf G}^{\cdot,\cdot}_{m,k}$ и
${\bf Q}^{\cdot\cdot,\cdot}_{m,k}$ были, соответст\-венно,
стаци\-онарными или периодичными по времени решениями задач (19)-(24).
Существо\-вание таких стационарных решений рассмотрено Желиговским [2003];
периоди\-ческий по времени случай рассматривается аналогично (см. Zheligovsky
и Podvigina [2005]).

Правые части уравнений (19)-(23) имеют нулевые средние, т.к.
исходные МГД состояния симметричны относительно начала координат, а функции
${\bf S}^{\cdot,\cdot}_k$ соответ\-ствуют центрально-антисиммет\-ричным состояниям.
В уравнении (4), записанном для вспомогательных задач (19) и (20) второго типа,
согласно условиям (19.2), (19.4), (20.2) и (20.4), а также в силу того,
что исходные поля $\bf V$ и $\bf H$ имеют симметрию, противоположную симметрии
полей ${\bf S}^{\cdot,\cdot}_k$, $\lad{\bf V}\nabla_{\bf x}\cdot{\bf G}^{\cdot,v}_{m,k}\rad=$\break
$\lad{\bf H}\nabla_{\bf x}\cdot{\bf G}^{\cdot,h}_{m,k}\rad=0$.
В уравнении (4), записанном для задачи (25),
$\lad{\bf V}\nabla_{\bf x}\cdot\bxi_1^v\rad=\lad{\bf H}\nabla_{\bf x}\cdot\bxi_1^h\rad=0$
согласно условиям (25.2) и (25.4) и поскольку $\nabla_{\bf x}\cdot\bxi_0^\cdot$
экспоненциаль\-но затухают во времени. Тем самым для всех рассматриваемых задач
(19)-(24) выполнено условие разрешимости (5), и они имеют единственные решения
соглас\-но предположению о разрешимости задач (6).

Поскольку пространства центрально-симметричных и центрально-антисим\-метричных
полей инвариантны для оператора ${\cal L}=({\cal L}^v,{\cal L}^h)$, решения ${\bf G}^{\cdot,\cdot}$
и $\bf Q^{\cdot\cdot,\cdot}$ соответствуют центрально-сим\-метричным состояниям:
$${\bf G}^{\cdot,v}_{m,k}(-{\bf x},t)=-{\bf G}^{\cdot,v}_{m,k}({\bf x},t),\quad
{\bf G}^{\cdot,h}_{m,k}(-{\bf x},t)=-{\bf G}^{\cdot,h}_{m,k}({\bf x},t),$$
$$G^{\cdot,p}_{m,k}(-{\bf x},t)=G^{\cdot,p}_{m,k}({\bf x},t);$$
$${\bf Q}^{\cdot\cdot,v}_{m,k}(-{\bf x},t)=-{\bf Q}^{\cdot\cdot,v}_{m,k}({\bf x},t),\quad
{\bf Q}^{\cdot\cdot,h}_{m,k}(-{\bf x},t)=-{\bf Q}^{\cdot\cdot,h}_{m,k}({\bf x},t),$$
$$Q^{\cdot\cdot,p}_{m,k}(-{\bf x},t)=Q^{\cdot\cdot,p}_{m,k}({\bf x},t).$$

Покажем, что решения задачи (25) экспоненциально затухают во времени.
(Аналогично доказывается, что экспоненциально затухают во времени изменения
функций ${\bf G}^{\cdot,\cdot}_{m,k}$ и ${\bf Q}^{\cdot\cdot,\cdot}_{m,k}$,
вызванные изменениями начальных условий для ${\bf S}^{\cdot,\cdot}_k$.)
Сделаем подстановку $\bxi_1^v=\hxi^v+\nabla_{\bf x}\xi^v$,
$\bxi_1^h=\hxi^h+\nabla_{\bf x}\xi^h$, где $\hxi^v$ и $\hxi^h$
соленоидальны, а $\xi$ -- глобально ограниченные вместе с производными
гладкие решения уравнений $\Delta_{\bf x}\xi^v+\nabla_{\bf X}\cdot\bxi_0^v=0$
и $\Delta_{\bf x}\xi^h+\nabla_{\bf X}\cdot\bxi_0^h=0$,
пространственные средние которых равны нулю (они тем самым также
экспоненциально затухают во времени). Обозначим ${\bf R(x},t)$ правую
часть уравнений (25.1), (25.3) после указанной подстановки (зави\-симость
функций от медленных переменных для простоты явно не указываем).

Усредняя уравнение ${\cal L}(\hxi)=\bf R$ с учетом (3), получаем
$$-{\partial\la\hxi\ra\over\partial t}=\la{\bf R}\ra\quad\Leftrightarrow\quad
-\la\hxi\ra|_{t=t_1}+\la\hxi\ra|_{t=0}=\int_0^{t_1}\la{\bf R}\ra dt.$$
Условие $\lad\hxi\rad=0$ влечет
$$\la\hxi\ra|_{t=0}=\lim_{\phantom{|}\tau\to\infty}{1\over\tau}\int_0^\tau
\int_0^{t_1}\la{\bf R}\ra\,dt\,dt_1\quad\Rightarrow$$
$$\la\hxi\ra|_{t=t_2}=\lim_{\phantom{|}\tau\to\infty}{1\over\tau}\int_0^\tau
\int_{t_2}^{t_1}\la{\bf R}\ra\,dt\,dt_1
=\lim_{\phantom{|}\tau\to\infty}{1\over\tau}\int_{t_2}^\tau
\int_{t_2}^{t_1}\la{\bf R}\ra\,dt\,dt_1.$$
Используя оценку
$|\la{\bf R(x},t)\ra|\le c_{\bf R}e^{-\alpha t}|{\bf R(x},0)|\ \ \forall t\ge0$,
находим отсюда
$$|\la\hxi({\bf x},t_2)\ra|\le(c_{\bf R}/\alpha)|{\bf R(x},0)|e^{-\alpha t_2}\ \ \forall t_2\ge0.$$

Таким образом, нам осталось показать, что если у решения уравнения\break
${\cal L}(\hat{\bxi})=\bf R$ компоненты $\hat{\bxi}^v$ и $\hat{\bxi}^h$ соленоидальны,
при $t=0$ их пространственные средние равны нулю, $\la{\bf R}\ra=0$,
и для нормы $\|\cdot\|$ имеет место оценка
$$\|{\bf R(x},t)\|\le C_{\bf R}e^{-\alpha t}\|{\bf R(x},0)\|\ \forall t\ge0,$$
то $\hat{\bxi}$ экспоненциально затухает в норме $\|\cdot\|$.

В соответствии с нашим предположением об исследуемом МГД состоянии, любое
глобально ограниченное решение $\bxi=(\bxi^v,\bxi^h,\xi^p)$ системы уравнений
$${\cal L}^v(\bxi^v,\bxi^h,\xi^p)=0,\quad{\cal L}^h(\bxi^v,\bxi^h)=0,\quad
\nabla_{\bf x}\cdot\bxi^v=\nabla_{\bf x}\cdot\bxi^h=0\eqn{26}$$
с любыми соленоидальными начальными данными с нулевыми средними
экспо\-ненциально затухает, т.е. для некоторых констант $C$ и $\alpha>0$
при любых \hbox{$t_1>t_2\ge 0$} выполнено неравенство
$$\|\bxi({\bf x},t_1)\|\le Ce^{-\alpha(t_1-t_2)}\|\bxi({\bf x},t_2)\|.\eqn{27}$$
(Без потери общности можно считать, что в (27) такой же показатель
экспоненты, как и в оценке для $\|{\bf R})\|$.)

Решение рассматриваемой задачи представим в виде суммы\break$\hxi=\bxi_I+\bxi_{II}$, где
$\bxi_I$ -- решение задачи (26) с неоднородными начальными данными
$\bxi_I|_{t=0}=\hat{\bxi}_1({\bf x},0)$, а $\bxi_{II}$ -- решение задачи
$${\cal L}^v(\bxi_{II}^v,\bxi_{II}^h,\xi_{II}^p)={\bf R}^v({\bf x},t),\quad
{\cal L}^h(\bxi_{II}^v,\bxi_{II}^h,)={\bf R}^h({\bf x},t),$$
$$\nabla_{\bf x}\cdot\bxi_{II}^v=\nabla_{\bf x}\cdot\bxi_{II}^h=0$$
с однородными начальными данными $\bxi_{II}|_{t=0}=0$. Согласно (27)
$$\|\bxi_I({\bf x},t)\|\le Ce^{-\alpha t}\|\bxi_I({\bf x},0)\|.\eqn{28}$$
По принципу Дюамеля
$$\bxi_{II}=\int_0^t\bxi_D({\bf x},t,\tau)d\tau,$$
где $\bxi_D({\bf x},t,\tau)$ при $t>\tau$ -- решение задачи (26)
c начальными данными\break$\bxi_D({\bf x},t,\tau)|_{t=\tau}={\bf R(x},\tau)$, и, значит,
для любого $\alpha'<\alpha$ имеют место оценки
$$\|\bxi_D({\bf x},t)\|\le Ce^{-\alpha(t-\tau)}C_{\bf R}e^{-\alpha\tau}\|{\bf R(x},0)\|
=CC_{\bf R}e^{-\alpha t}\|{\bf R(x},0)\|$$
$$\Rightarrow\quad\|\bxi_{II}({\bf x},t)\|\le CC_{\bf R}te^{-\alpha t}\|{\bf R(x},0)\|
\le C_{\alpha'}e^{-\alpha't}\|{\bf R(x},0)\|.$$
Это неравенство совместно с (28) влечет желаемый результат.

\mi{\bf 6. Уравнения порядка $\varepsilon^2$.}
Уравнения, полученные из членов рядов (2.1) и (2.2) порядка $\varepsilon^2$,
имеют вид

\pagebreak
$${\cal L}^v(\lbd{\bf v}_2\rbd,\lbd{\bf h}_2\rbd,\lbd p_2\rbd)
-{\partial{\bf v}_0\over\partial T}+\nu(2(\nabla_{\bf x}\cdot\nabla_{\bf X})\lbd{\bf v}_1\rbd+\Delta_{\bf X}{\bf v}_0)
+\lad{\bf v}_2\rad\times(\nabla_{\bf x}\times{\bf V})$$
$$+(\nabla_{\bf x}\times{\bf H})\times\lad{\bf h}_2\rad
+{\bf V}\times(\nabla_{\bf X}\times{\bf v}_1)
+(\nabla_{\bf X}\times{\bf h}_1)\times{\bf H}$$
$$+{\bf v}_0\times(\nabla_{\bf x}\times\lbd{\bf v}_1\rbd)
+{\bf v}_1\times(\nabla_{\bf x}\times\lbd{\bf v}_0\rbd)
+{\bf v}_0\times(\nabla_{\bf X}\times{\bf v}_0)$$
$$+(\nabla_{\bf x}\times\lbd{\bf h}_0\rbd)\times{\bf h}_1
+(\nabla_{\bf x}\times\lbd{\bf h}_1\rbd)\times{\bf h}_0
+(\nabla_{\bf X}\times{\bf h}_0)\times{\bf h}_0-\nabla_{\bf X}p_1=0,$$
$${\cal L}^h(\lbd{\bf v}_2\rbd,\lbd{\bf h}_2\rbd)-{\partial{\bf h}_0\over\partial T}
+\eta\left(2(\nabla_{\bf x}\cdot\nabla_{\bf X})\lbd{\bf h}_0\rbd+\Delta_{\bf X}{\bf h}_0\right)
+(\lad{\bf h}_2\rad\cdot\nabla_{\bf x}){\bf V}$$
$$-(\lad{\bf v}_2\rad\cdot\nabla_{\bf x}){\bf H}
+\nabla_{\bf X}\times({\bf v}_1\times{\bf H}+{\bf V}\times{\bf h}_1+{\bf v}_0\times{\bf h}_0)
+\nabla_{\bf x}\times({\bf v}_1\times{\bf h}_0+{\bf v}_0\times{\bf h}_1)=0.$$
Их средние части (использовано (10) при $n=1,2$)
$$-{\partial\over\partial T}\lad{\bf v}_0\rad
+\nu\nabla^2_{\bf X}\lad{\bf v}_0\rad-\sum_{j=1}^3{\partial\over\partial X_j}\ladb
{\bf V}^j\lbd{\bf v}_1\rbd+{\bf V}\lbd{\bf v}_1^j\rbd+{\bf v}_0^j{\bf v}_0$$
$$-{\bf H}^j\lbd{\bf h}_1\rbd-{\bf H}\lbd{\bf h}_1^j\rbd-{\bf h}_0^j{\bf h}_0\radb
-\nabla_{\bf X}p'=0,$$
где $p'({\bf X},T)=\lad p_1-(|{\bf v}_0|^2-|{\bf h}_0|^2)/2-{\bf V}\cdot{\bf v}_1+{\bf H}\cdot{\bf h}_1\rad$, и
$$-{\partial\over\partial T}\lad{\bf h}_0\rad+\eta\Delta_{\bf X}\lad{\bf h}_0\rad
+\nabla_{\bf X}\times\lad\lbd{\bf v}_1\rbd\times{\bf H}+{\bf V}\times\lbd{\bf h}_1\rbd
+{\bf v}_0\times{\bf h}_0\rad=0$$
в силу (12) и (18) принимают вид
$$-{\partial\over\partial T}\lad{\bf v}_0\rad
+\nu\Delta_{\bf X}\lad{\bf v}_0\rad+\sum_{j=1}^3\sum_{m=1}^3\sum_{k=1}^3
\left({\bf D}^{v,v}_{m,k,j}{\partial^2\lad{\bf v}_0^k\rad\over\partial X_j\partial X_m}
+{\bf D}^{h,v}_{m,k,j}{\partial^2\lad{\bf h}_0^k\rad\over\partial X_j\partial X_m}\right)$$
$$+\sum_{j=1}^3\sum_{m=1}^3\sum_{k=1}^3{\partial\over\partial X_j}\left(
{\bf A}^{vv,v}_{m,k,j}\lad{\bf v}^k_0\rad\lad{\bf v}^m_0\rad
+{\bf A}^{vh,v}_{m,k,j}\lad{\bf v}^k_0\rad\lad{\bf h}^m_0\rad
+{\bf A}^{hh,v}_{m,k,j}\lad{\bf h}^k_0\rad\lad{\bf h}^m_0\rad\right)$$
$$+(\lad{\bf v}_0\rad\cdot\nabla_{\bf X})\lad{\bf v}_0\rad
-(\lad{\bf h}_0\rad\cdot\nabla_{\bf X})\lad{\bf h}_0\rad
-\nabla_{\bf X}p'=0,\eqn{29.1}$$
$$-{\partial\over\partial T}\lad{\bf h}_0\rad+\eta\Delta_{\bf X}\lad{\bf h}_0\rad
+\nabla_{\bf X}\times\left(\sum_{m=1}^3\sum_{k=1}^3
\left({\bf D}^{v,h}_{m,k}{\partial\lad{\bf v}_0^k\rad\over\partial X_m}
+{\bf D}^{h,h}_{m,k}{\partial\lad{\bf h}_0^k\rad\over\partial X_m}\right)\right.
+\lad{\bf v}_0\rad\times\lad{\bf h}_0\rad$$
$$+\left.\sum_{m=1}^3\sum_{k=1}^3\left(
{\bf A}^{vv,h}_{m,k}\lad{\bf v}^k_0\rad\lad{\bf v}^m_0\rad
+{\bf A}^{vh,h}_{m,k}\lad{\bf v}^k_0\rad\lad{\bf h}^m_0\rad
+{\bf A}^{hh,h}_{m,k}\lad{\bf h}^k_0\rad\lad{\bf h}^m_0\rad\right)\right)=0.\eqn{29.2}$$
Здесь $\bf D$ обозначают коэффициенты членов уравнений средних полей, отвечающие
т.н. вихревой диффузии, $\bf A$ -- квадратичных членов т.н. вихревой адвекции:
$${\bf D}^{v,v}_{m,k,j}=\lad
-{\bf V}^j{\bf G}^{v,v}_{m,k}-{\bf V}({\bf G}^{v,v}_{m,k})^j
+{\bf H}^j{\bf G}^{v,h}_{m,k}+{\bf H}({\bf G}^{v,h}_{m,k})^j\rad,\eqn{30.1}$$
$${\bf D}^{h,v}_{m,k,j}=\lad
-{\bf V}^j{\bf G}^{h,v}_{m,k}-{\bf V}({\bf G}^{h,v}_{m,k})^j
+{\bf H}^j{\bf G}^{h,h}_{m,k}+{\bf H}({\bf G}^{h,h}_{m,k})^j\rad,\eqn{30.2}$$
$${\bf D}^{v,h}_{m,k}=\lad{\bf V}\times{\bf G}^{v,h}_{m,k}-{\bf H}\times{\bf G}^{v,v}_{m,k}\rad,\eqn{30.3}$$
$${\bf D}^{h,h}_{m,k}=\lad{\bf V}\times{\bf G}^{h,h}_{m,k}-{\bf H}\times{\bf G}^{h,v}_{m,k}\rad\eqn{30.4}$$
(коэффициенты (30) идентичны коэффициентам тензора комбинированной вихре\-вой
диффузии в задаче о линейной устойчивости МГД стационарных состояний [Желиговский, 2003]),
$${\bf A}^{vv,v}_{m,k,j}=\lad
-{\bf V}^j{\bf Q}^{vv,v}_{m,k}-{\bf V}({\bf Q}^{vv,v}_{m,k})^j
+{\bf H}^j{\bf Q}^{vv,h}_{m,k}+{\bf H}({\bf Q}^{vv,h}_{m,k})^j
+({\bf S}^{v,v}_k)^j{\bf S}^{v,v}_m-({\bf S}^{v,h}_k)^j{\bf S}^{v,h}_m\rad,\eqn{31.1}$$
$${\bf A}^{vh,v}_{m,k,j}=\lad
-{\bf V}^j{\bf Q}^{vh,v}_{m,k}-{\bf V}({\bf Q}^{vh,v}_{m,k})^j
+{\bf H}^j{\bf Q}^{vh,h}_{m,k}+{\bf H}({\bf Q}^{vh,h}_{m,k})^j$$
$$+({\bf S}^{v,v}_k)^j{\bf S}^{h,v}_m+({\bf S}^{h,v}_m)^j{\bf S}^{v,v}_k
-({\bf S}^{v,h}_k)^j{\bf S}^{v,h}_m-({\bf S}^{h,h}_m)^j{\bf S}^{v,h}_k\rad,\eqn{31.2}$$
$${\bf A}^{hh,v}_{m,k,j}=\lad
-{\bf V}^j{\bf Q}^{hh,v}_{m,k}-{\bf V}({\bf Q}^{hh,v}_{m,k})^j
+{\bf H}^j{\bf Q}^{hh,h}_{m,k}+{\bf H}({\bf Q}^{hh,h}_{m,k})^j
+({\bf S}^{h,v}_k)^j{\bf S}^{h,v}_m-({\bf S}^{h,h}_k)^j{\bf S}^{h,h}_m\rad,\eqn{31.3}$$
$${\bf A}^{vv,h}_{m,k}=\lad{\bf V}\times{\bf Q}^{vv,h}_{m,k}-{\bf H}\times{\bf Q}^{vv,v}_{m,k}
+{\bf S}^{v,v}_k\times{\bf S}^{v,h}_m\rad,\eqn{31.4}$$
$${\bf A}^{vh,h}_{m,k}=\lad{\bf V}\times{\bf Q}^{vh,h}_{m,k}-{\bf H}\times{\bf Q}^{vh,v}_{m,k}
+{\bf S}^{v,v}_k\times{\bf S}^{h,h}_m+{\bf S}^{h,v}_m\times{\bf S}^{v,h}_k\rad,\eqn{31.5}$$
$${\bf A}^{hh,h}_{m,k}=\lad{\bf V}\times{\bf Q}^{hh,h}_{m,k}-{\bf H}\times{\bf Q}^{hh,v}_{m,k}
+{\bf S}^{h,v}_k\times{\bf S}^{h,h}_m\rad.\eqn{31.6}$$

\mi{\bf 7. Вычисление коэффициентов вихревых членов.}
Для вычисления коэф\-фициентов $\bf A$ и $\bf D$ в уравнениях (29) достаточно
решить 45 вспомогательных задач (6 первого, 18 второго и 21 третьего типа;
формально уравнения (21)-(24) определяют 27 задач третьего типа, однако
в коэффициенты (30) и (31) поля ${\bf Q}^{vv,\cdot}_{m,k}$
и ${\bf Q}^{hh,\cdot}_{m,k}$ входят не индивидуально, а в виде сумм
${\bf Q}^{vv,\cdot}_{m,k}+{\bf Q}^{vv,\cdot}_{k,m}$ и
${\bf Q}^{hh,\cdot}_{m,k}+{\bf Q}^{hh,\cdot}_{k,m}$, и, соответственно,
можно уменьшить число решаемых задач (21) и (23) в обоих случаях на 3,
если их при $m\ne k$ переформулировать для этих сумм). Число вспомогательных
задач, которые необходимо решить, можно уменьшить втрое, если ввести в рассмотрение
(как в работе [Zheligovsky, 2005]) {\it вспомогательные задачи для сопряженного
оператора} (при этом вычислительная сложность реша\-емых задач не увеличивается):
$$({\cal L}^*)^v({\bf Z}^{v,v}_{j,n},{\bf Z}^{v,h}_{j,n})={\cal P}(-{\bf V}^j{\bf e}_n-{\bf V}^n{\bf e}_j),\eqn{32.1}$$
$$({\cal L}^*)^h({\bf Z}^{v,v}_{j,n},{\bf Z}^{v,h}_{j,n})={\cal P}({\bf H}^j{\bf e}_n+{\bf H}^n{\bf e}_j),\eqn{32.2}$$
$$\nabla\cdot{\bf Z}^{v,v}_{j,n}=\nabla\cdot{\bf Z}^{v,h}_{j,n}=0;\eqn{32.3}$$
$$({\cal L}^*)^v({\bf Z}^{h,v}_{j,n},{\bf Z}^{h,h}_{j,n})={\cal P}({\bf H}\times{\bf e}_n),\eqn{33.1}$$
$$({\cal L}^*)^h({\bf Z}^{h,v}_{j,n},{\bf Z}^{h,h}_{j,n})=-{\cal P}({\bf V}\times{\bf e}_n),\eqn{33.2}$$
$$\nabla\cdot{\bf Z}^{h,v}_{j,n}=\nabla\cdot{\bf Z}^{h,h}_{j,n}=0.\eqn{33.3}$$
Здесь ${\cal L}^*=(({\cal L}^*)^v,\,({\cal L}^*)^h)$ -- оператор, сопряженный к ${\cal L}$,
формально определенный в пространстве пар трехмерных соленоидальных полей:
$$({\cal L}^*)^v({\bf v,h})={\partial{\bf v}\over\partial t}+\nu\Delta{\bf v}
-\nabla\times({\bf V}\times{\bf v})+{\cal P}({\bf H}\times(\nabla\times{\bf h})
-{\bf v}\times(\nabla\times{\bf V})),$$
$$({\cal L}^*)^h({\bf v,h})={\partial{\bf h}\over\partial t}+\eta\Delta{\bf h}
+\nabla\times({\bf H}\times{\bf v})+{\cal P}({\bf v}\times(\nabla\times{\bf H})
-{\bf V}\times(\nabla\times{\bf h}))$$
(все дифференциальные операторы -- в быстрых переменных), $\cal P$ -- оператор
проекции трехмерного векторного поля в пространст\-во соленоидальных полей.

Все средние от произведений решений задач (19)-(24) с полями $\bf V$ и $\bf H$,
входя\-щие в (30) и (31), выражаются через решения вспомогательных задач (32)\break
(6 задач) и (33) (3 задачи) и правые части уравнений (19)-(23) равенствами
$$\lad-{\bf V}^j{\bf q}^v-{\bf V}({\bf q}^v)^j+{\bf H}^j{\bf q}^h+{\bf H}({\bf q}^h)^j\rad^n$$
$$=\lad-({\cal I}-{\cal P})({\bf V}^j{\bf e}_n-{\bf V}^n{\bf e}_j)\cdot({\cal I}-{\cal P}){\bf q}^v
+({\cal I}-{\cal P})({\bf H}^j{\bf e}_n-{\bf H}^n{\bf e}_j)\cdot({\cal I}-{\cal P}){\bf q}^h\rad$$
$$+\lad({\cal L}^*)^v({\bf Z}^{v,v}_{j,n},{\bf Z}^{v,h}_{j,n})\cdot{\cal P}{\bf q}^v
+({\cal L}^*)^h({\bf Z}^{v,v}_{j,n},{\bf Z}^{v,h}_{j,n})\cdot{\cal P}{\bf q}^h\rad$$
$$=\lad-({\cal I}-{\cal P})({\bf V}^j{\bf e}_n-{\bf V}^n{\bf e}_j)\cdot({\cal I}-{\cal P}){\bf q}^v
+({\cal I}-{\cal P})({\bf H}^j{\bf e}_n-{\bf H}^n{\bf e}_j)\cdot({\cal I}-{\cal P}){\bf q}^h\rad$$
$$+\lad{\bf Z}^{v,v}_{j,n}\cdot{\cal L}^v({\cal P}{\bf q}^v,{\cal P}{\bf q}^h)
+{\bf Z}^{v,h}_{j,n}\cdot{\cal L}^h({\cal P}{\bf q}^v,{\cal P}{\bf q}^h)\rad;\eqn{34}$$
$$\lad{\bf V}\times{\bf q}^h-{\bf H}\times{\bf q}^v\rad^n
=\lad{\bf q}^v\cdot({\bf H}\times{\bf e}_n)-{\bf q}^h\cdot({\bf V}\times{\bf e}_n)\rad$$
$$=\lad({\cal I}-{\cal P}){\bf q}^v\cdot({\cal I}-{\cal P})({\bf H}\times{\bf e}_n)
-({\cal I}-{\cal P}){\bf q}^h\cdot({\cal I}-{\cal P})({\bf V}\times{\bf e}_n)\rad$$
$$+\lad{\cal P}{\bf q}^v\cdot({\cal L}^*)^v({\bf Z}^{h,v}_n,{\bf Z}^{h,h}_n)
+{\cal P}{\bf q}^h\cdot({\cal L}^*)^h({\bf Z}^{h,v}_n,{\bf Z}^{h,h}_n)\rad$$
$$=\lad({\cal I}-{\cal P}){\bf q}^v\cdot({\cal I}-{\cal P})({\bf H}\times{\bf e}_n)
-({\cal I}-{\cal P}){\bf q}^h\cdot({\cal I}-{\cal P})({\bf V}\times{\bf e}_n)\rad$$
$$+\lad{\cal L}^v({\cal P}{\bf q}^v,{\cal P}{\bf q}^h)\cdot{\bf Z}^{h,v}_n
+{\cal L}^h({\cal P}{\bf q}^v,{\cal P}{\bf q}^h)\cdot{\bf Z}^{h,h}_n)\rad.\eqn{35}$$
Здесь $\cal I$ -- тождественный оператор, ${\bf q}^v,{\bf q}^h$ --
решение какой-либо из задач (19), (20) при подсчете коэффициентов $\bf D$
и задач (21)-(24) при подсчете коэффициен\-тов $\bf A$. Для задач (19) и (20)
${\cal P}{\bf q}^v,{\cal P}{\bf q}^h$ вычисляется из соотношений (19.2), (19.4), (20.2) и
(20.4), для задач (21)-(24) ${\cal P}{\bf q}^v={\cal P}{\bf q}^h=0$. (Как обычно,
вычисление проекции $\cal P$ удобно делать с помощью дискретного или непрерывного
преобразо\-вания Фурье.) Тем самым, решение вспомогательных задач второго
и третьего типов оказывается возможным избежать.

Формально сопряженный оператор ${\cal L}^*$ определен корректно,
например, если исходное состояние ${\bf V,H},P$ периодично по прост\-ранству
и стационарно или пери\-одично по времени, и рассматривается область определения
оператора ${\cal L}^*$, состо\-ящая из функций, имеющих такие же свойства стационарности
и/или периодич\-ности. В общем случае численное решение задач (32) и (33)
проблематично, т.к. оператор ${\cal L}^*$ не параболический. Обойти эту сложность можно
следующим образом: для неизвестных полей ${\bf Z}^{\cdot,\cdot}$ нулевые
``начальные" условия ставим при $t=\tau>0$, а затем уравнения (32) и (33) решаем
по времени ``назад", в сторону убывающего $t$ до $t=0$ (при обращении времени
операторы $({\cal L}^*)^v,\,({\cal L}^*)^h$ становятся параболическими). Полученное решение,
зависящее от $\tau$, обозначим ${\bf Z}^{\cdot,\cdot}(\tau;{\bf x},t)$. Тогда
пространственно-временн\'ое усреднение по быстрым переменным скалярных
произведений с функциями ${\bf Z}^{\cdot,\cdot}$ в (34) и (35) определяется равенством
$$\lad{\bf f}\cdot{\bf Z}^{\cdot,\cdot}\rad=\lim_{\phantom{\ell}\tau\to\infty}
\lim_{\phantom{\ell}\ell\to\infty}{1\over\tau\ell^3}\int_0^\tau\int_{-\ell/2}^{\ell/2}
\int_{-\ell/2}^{\ell/2}\int_{-\ell/2}^{\ell/2}
{\bf f}({\bf x},t){\bf Z}^{\cdot,\cdot}(\tau;{\bf x},t)\,d{\bf x}\,dt$$
(если указанный предел существует, -- например, если
исходное состояние ${\bf V,H},P$ квазипериодично по времени).

\mi{\bf 8. Выводы.}
Рассмотрена устойчивость центрально-симметрич\-ного МГД состоя\-ния,
не имеющего больших масштабов, по отношению к возмущению, в котором
присутствуют большие прост\-ранственные и временн\'ые масштабы.
Построено асимптотическое разложение возмущения и выведены уравнения (29)
эволюции в нелинейном режиме главного члена разложения возмущения, усредненного по малым
пространственно-временн\'ым масштабам. Предложен метод экономичного вычисления
коэффициентов вихревой диффузии (30) и адвекции (31) в уравнени\-ях для средних
полей (29), использующий выражение этих коэффициентов через решения
вспомогательных задач для оператора, сопряженного к оператору лине\-аризации
исходной системы уравнений магнитогидродинамики в окрестности МГД состояния,
нелинейная устойчивость которого исследуется.

\mi{\bf Благодарности}.
Работа частично финансировалась РФФИ (грант 04-05-64699).

\mi{\bf Литература}

\mi{\it В.А.Желиговский.} О линейной устойчивости стационарных
прост\-ранственно-периодических магнитогидродинамических систем
к длиннопериодным возмуще\-ниям // Физика Земли, 2003. $N$ 5. C. 65-74
[http://arxiv.org/abs/nlin/0512076].

\mi{\it Baptista M., Gama S., Zheligovsky V.}
Multiple-scale expansions on incompressible MHD systems.
Preprint 2004-11, Centro de Matem\'atica da Universidade do Porto,
Faculdade de Ci\^encias da Universidade do Porto\break
[http://cmup.fc.up.pt/cmup/preprints/2004-11.pdf].

\mi{\it Bensoussan A., Lions J.-L., Papanicolaou G.} Asymptotic
Analysis for Periodic Structures. North Holland, 1978.


\mi{\it Cioranescu D., Donato P.} An introduction to homogenization. Oxford
Univ. Press, 1999.

\mi{\it Gama S., Vergassola M., Frisch U.} Negative eddy viscosity in
isotropically forced two-dimensional flow: linear and nonlinear dynamics //
J. Fluid Mech., 1994. V. 260, P. 95-126.

\mi{\it Newell A.C., Passot T., Lega J.} Order parameter equations for patterns
// Ann. Rev. Fluid Mech., 1993. V. 50, P. 399-453.

\mi{\it Oleinik O.A., Shamaev A.S., Yosifian G.A.} Mathematical problems
in elasticity and homogenization. Amsterdam, Elsevier Science Publishers, 1992.

\mi{\it Zheligovsky V.A., Podvigina O.M., Frisch U.} Dynamo effect
in parity-invariant flow with large and moderate separation of scales //
Geophys. Astrophys. Fluid Dynamics, 2001. V. 95. P. 227-268
[http://xxx.lanl.gov/abs/nlin.CD/0012005].

\mi{\it Zheligovsky V.A., Podvigina O.M.} Generation of multiscale magnetic field
by parity-invariant time-periodic flows. Geophys. Astrophys. Fluid
Dynamics, 2003. V. 97. P. 225-248 [http://xxx.lanl.gov/abs/physics/0207112].

\mi{\it Zheligovsky V.A.} Convective plan-form two-scale dynamos in a plane layer.
Geophys. Astrophys. Fluid Dynamics, 2005. V. 99. P. 151-175\break
[http://arxiv.org/abs/physics/0405045].
\end{document}